\documentclass[%
 aip,
 amsmath,amssymb,
 reprint,%
]{revtex4-1}
\usepackage{multirow}
\usepackage{mathrsfs}  
\usepackage[section]{placeins}
\usepackage{graphicx}
\usepackage{dcolumn}
\usepackage{bm}

\usepackage[utf8]{inputenc}
\usepackage[T1]{fontenc}
\usepackage{mathptmx}
\usepackage{etoolbox}
\usepackage[normalem]{ulem}
\usepackage[]{xcolor}

\makeatletter
\def\@email#1#2{%
 \endgroup
 \patchcmd{\titleblock@produce}
  {\frontmatter@RRAPformat}
  {\frontmatter@RRAPformat{\produce@RRAP{*#1\href{mailto:#2}{#2}}}\frontmatter@RRAPformat}
  {}{}
}%

\makeatother
\begin{document}


\title[]{Tunable turbulence in driven microscale emulsions}
\author{Majid Bahraminasr}
\altaffiliation[Also at ]{Department of Physics, University of Guelph, Guelph, Canada}
\affiliation{
Department of Physics and Physical Oceanography, Memorial University, St. John’s, NL A1B 3X7, Canada
}%

\author{Anand Yethiraj}%
\homepage{https://softmaterials.ca/}
 \email{ayethira@uoguelph.ca}
\affiliation{Department of Physics, University of Guelph, Guelph, ON N1G 2W1, Canada
}%

\newcommand{\aycomment}[1]{{\color{red}[\bf{#1}]}}
\newcommand{\mb}[1]{{\color{brown}{#1}}}
\newcommand{\mbcomment}[1]{{\color{green}[\bf{#1}]}}

\date{\today}

\begin{abstract}
 We present a tunable, non-equilibrium oil-in-oil emulsion that serves as a model system for investigating the transition from controlled droplet deformation to  multiscale flows reminiscent of turbulence. By utilizing a miscible mixture of silicone and motor oils as the continuous phase and the immiscible castor oil as the droplet phase, we isolate electrical conductivity as a single experimental control parameter, varying it by over two orders of magnitude while keeping viscosity and permittivity nearly constant. This high degree of control allows us to systematically traverse the electrohydrodynamic (EHD) phase diagram with dielectric constant and conductivity as control parameters. We validate small-deformation theory at low fields before driving the system into a regime of multiscale, unsteady flows at high fields. We employ three complementary approaches on the same system (particle image velocimetry (PIV), used to map velocity fields, and rheometry and differential dynamic microscopy (DDM), two techniques used to probe viscosity and diffusion) to quantify the emergence of scale invariance in the energy spectra with increasing field strength. Above a threshold field, we find that the spatio-temporal energy spectra obtained by PIV analysis of droplet dynamics display power-law scaling, $E(k) \sim k^{-\alpha_k}$, where $\alpha_k$ approaches the inertial turbulence exponent of $5/3$ at high fields. Energy spectra from rheometry also yield a power law, $S(\nu) \sim \nu^{-\alpha_\nu}$, with $\alpha_\nu = 5/3$ at high fields. Mean square displacement (MSD) analyses on the same datasets reveal super-diffusive behavior, $\mathrm{MSD} \sim t^{\gamma}$, with $\gamma = 3/2$. These observations provide strong evidence of a conductivity-tunable transition to EHD-driven turbulence in a microscale emulsion.

\end{abstract}

\maketitle

\section{\label{sec:intro}Introduction}

The manipulation of droplets under electric fields has been a cornerstone of electrohydrodynamics (EHD), with implications ranging from lab-on-a-chip technologies \cite{zeng2004principles, lee2002electrowetting, mugele2005electrowetting, liu2023electrowetting, Varshney2014} to the nucleation and growth of crystals \cite{sanz2007evidence,langer1980instabilities,lowen2004marriage}.
Taylor first reported the phenomenon in 1966, introduced the leaky dielectric model (LDM) to account for the role of non-zero conductivity on the electric response, and predicted the steady deformation of single-phase droplets under an electric field~\cite{taylor1966studies, melcher1969electrohydrodynamics, saville1997electrohydrodynamics}. The LDM assumes that the interface carries no net charge and that the bulk liquids remain electroneutral 
i.e., it neglects charge convection. Subsequent studies~\cite{torza1971electrohydrodynamic} validated the model, 
while more refined theoretical and computational approaches~\cite{ajayi1978note, feng1996computational, bentenitis2005droplet, zhang2013transient, zholkovskij2002electrokinetic, moriya1986deformation, zabarankin2020small, esmaeeli2011transient} 
advanced the theory but were limited to the low-field, \emph{linear regime}, i.e. first-order small-deformation expansions in the electric capillary number $Ca_E$~\cite{taylor1966studies}. Second-order corrections~\cite{ajayi1978note} still neglected charge convection and transient dynamics. Later work emphasized that charge relaxation and nonlinear effects are critical for quantitative accuracy~\cite{lanauze2013influence, salipante2010electrohydrodynamics}, and have 
established a nonlinear small-deformation framework incorporating transient relaxation and charge convection~\cite{das2017nonlinear}.

Accounting for charge convection requires solving the coupled electrostatic and flow fields simultaneously~\cite{vlahovska2019electrohydrodynamics, das2017nonlinear}, which is challenging. The resulting nonlinear feedback between charge transport and fluid motion means that neglecting convection leads to systematic errors in predicting droplet deformation~\cite{feng1999electrohydrodynamic, supeene2008deformation, lopez2011charge, lanauze2015nonlinear, tomar2007two}, breakup dynamics~\cite{sengupta2017role}, and unsteady phenomena such as electrorotation~\cite{dong2018electrohydrodynamic}. Dong et al.~\cite{dong2018electrohydrodynamic} showed that outcomes are governed by competition between charge convection driven by fluid motion and charge supply through conduction. 
Experiments have used viscosity ratio as a tuning parameter because silicone oils are commercially available in multiple viscosities; however, varying the conductivity ratio would provide more direct control of the convection–conduction balance. 
Torza {\it et al.}~\cite{torza1971electrohydrodynamic}, identified three drop deformation regimes as a function of the dielectric constant and conductivity ratios. The simplest way to traverse this phase diagram is by varying the conductivity ratio. Such a direct probe of the phase diagram has not been attempted experimentally.

In this work, we utilize the miscible mixture of motor oil and silicone oil as the continuous phase, with castor oil as the dispersed phase.
A unique aspect of this system 
is that motor oil exhibits significantly higher conductivity than silicone oil, while being nearly perfectly matched in relative permittivity and viscosity. 
Silicone oil and motor oil mixtures are miscible, and allow us to create a continuous phase with fixed relative permittivity and viscosity, but with variable conductivity 
that allows us to systematically lower the threshold for field-induced droplet instabilities and access nonlinear, unsteady regimes at reduced voltages—regimes that are otherwise difficult to achieve with conventional silicone oil systems.  

We report multiscale flows that are strongly analogous to turbulent flows. Recognizing the high bar for making this analogy, we quantify the resulting dynamics adopting three complementary experimental approaches on the same sample: particle image velocimetry (PIV), the standard tool in hydrodynamics for measuring flow fields, and rheometry and differential dynamic microscopy (DDM), widely used in soft-matter physics to probe viscosity and diffusive/super-diffusive transport. These techniques are rarely combined in a single study, yet together they enable a unified characterization of both microscopic diffusion and macroscopic turbulence in electrohydrodynamic emulsions. We systematically tune the inverse conductivity ratio through controlled oil mixing and show that the electric-field threshold for these multiscale flows is not fixed but can be modulated.

\section{Theory and the parameter space for experiments}

The deformation of a droplet in the presence of electrohydrodynamic forcing is determined by a balance of three forces: the bound polarization charge (normal to the interface), the induced free charge (both normal and tangential to the interface), and viscous forces (also in both normal and tangential directions to the interface). The detailed dependencies as a function of system material parameters, field strength, and field frequency were obtained in the formative work of Torza, Cox and Mason~\cite{torza1971electrohydrodynamic}. 
The drop deformation parameter is defined as $D=\frac{d_{\parallel}-d_{\perp}}{d_{\parallel}+d_{\perp}}$, where $d_{\parallel}$ ($d_{\perp}$) represents the dimension of the deformed droplet parallel (perpendicular) to the field. 
When a drop undergoes prolate or oblate deformation, $D$ takes a positive ($D>0$) or negative ($D<0$) value, respectively, while $D=0$ for an undeformed spherical drop. The key material parameters in this system are the interfacial tension, $\gamma_{\text{int}}$, between the two fluids and the three ratios of conductivity ($H = \sigma_1/\sigma_2$), relative dielectric permittivity ($S = \epsilon_1/\epsilon_2$) and viscosity ($N = \eta_1/\eta_2$), where the subscript 1 and 2 denote the dispersed phase and continuous phase, respectively. 
The drop deformation as a function of frequency  $\nu$  
is given by~\cite{torza1971electrohydrodynamic}
\begin{equation}
D_{\nu}=\frac{9}{16}\Phi_\nu\frac{\epsilon_0\kappa_2 a\bar{E}_0^2}{\gamma_\text{int}},
\label{eq:deformation}
\end{equation} 
where $a$ is the drop radius, $\bar{E}_0$ is the electric field amplitude, $\tau_{21} = \epsilon_0 \kappa_2/\sigma_1$, and $\Phi_\nu(H,S,N,\omega)$ is a discriminant function defined as
\begin{equation*}
\Phi_\nu
= 1 -
\frac{
f_1(H,S,N)
  + 15 \tau_{21}^{2}\omega^{2}(1+N)(1+2S)
}{
  5(1+N)\!\left[\left(\frac{2}{H}+1\right)^{2} + \tau_{21}^{2}\omega^{2}(S+2)^{2}\right]
},
\label{eq:discriminationFunction}
\end{equation*}
where 
$f_1(H,S,N) = \frac{11N+14}{H}
  + \frac{15(N+1)+ S(19N+16)}{H^{2}}$, and $\omega=2\pi\nu$.
%
A critical frequency $\nu_c$, signifying a change in sign of the discrimination function, can be obtained by finding the root of Eq.~\ref{eq:discriminationFunction}, i.e., $\Phi_\nu = 0$, which yields
\begin{equation}
\begin{split}
\nu_c &=
\frac{1}{2\pi}\frac{
  \sqrt{
    \frac{S(19N + 16) - 5(1 + N)}{H^2}
    - \frac{3(3N + 2)}{H}
    - 5(1 + N)
  }
}{
  2\tau |S - 1| \sqrt{5(1 + N)}
}.
\end{split}
\label{eq:critical_freq}
\end{equation}

Depending on the value of the ratio $S/H$, three classes of deformation
are possible~\cite{torza1971electrohydrodynamic}. 
%
%
%
\begin{table}[h!]
\centering
\begin{ruledtabular}
\begin{tabular}{cccc}
 & Class A & Class B & Class C \\
\hline
$D_\nu$ & $>0$ & $>0$ & $\lesseqgtr 0\; (\nu \lesseqgtr \nu_c)$ \\
\hline
$\partial D_\nu/\partial \nu$ & $\leq 0$ & $>0$ & $>0$ \\
\end{tabular}
\end{ruledtabular}
\caption{Three classes of drop deformation $D_{\nu}$ as a function of frequency $\nu$. Class A drops are always prolate but become less prolate with increasing $\nu$. Class B has the smallest parameter space: here the drops are prolate but become more prolate with increasing $\nu$. Class C drops are oblate at low frequency, and there is a critical frequency $\nu_c$ above which they become prolate.}
\label{tab:Classes}
\end{table}
For class A systems~\cite{torza1971electrohydrodynamic}, $S/H < 1$:  $D_{\nu} > 0$, i.e., the drop is prolate (see Table~\ref{tab:Classes}). If $\frac{\partial D_\nu}{\partial \nu} < 0$, the drops become less prolate with increasing frequency, while when $\partial D_\nu/\partial \nu = 0$, the drop shape is independent of frequency. 
In a class B system, $1 < S/H <  1+\frac{5S(1+N)(1/S-1)^2}{16+19N}$, $D_{\nu} > 0$ (the drop deformation is prolate) and $\frac{\partial D_\nu}{\partial \nu} > 0$ (it becomes more prolate with increasing frequency). 
Above a critical field strength in class A and class B systems, the prolate droplets break up and disintegrate~\cite{nganguia2013equilibrium, zabarankin2013viscous, zhang2013transient, yariv2013electrohydrodynamic, lac2007axisymmetric, karyappa2014breakup, pillai2016electrokinetics}.
To our knowledge, class B systems have not been reported in experiment.
Finally, in Class C systems, 
$S/H >  1+\frac{5S(1+N)(1/S-1)^2}{16+19N}$. Here the deformation changes shape as the frequency crosses a critical value $\nu_c$: $D_{\nu} \lesseqgtr 0$ when $\nu \lesseqgtr \nu_c$. This change of sign {\it and} shape
serves as an experimental marker for identifying $\nu_c$.
In both class B and C, $\frac{\partial D_\nu}{\partial \nu}> 0$ means that 
the drop becomes more prolate with increasing frequency. 
%
%
\begin{figure*}[ht!]
\centering
\includegraphics[width=0.65\textwidth]{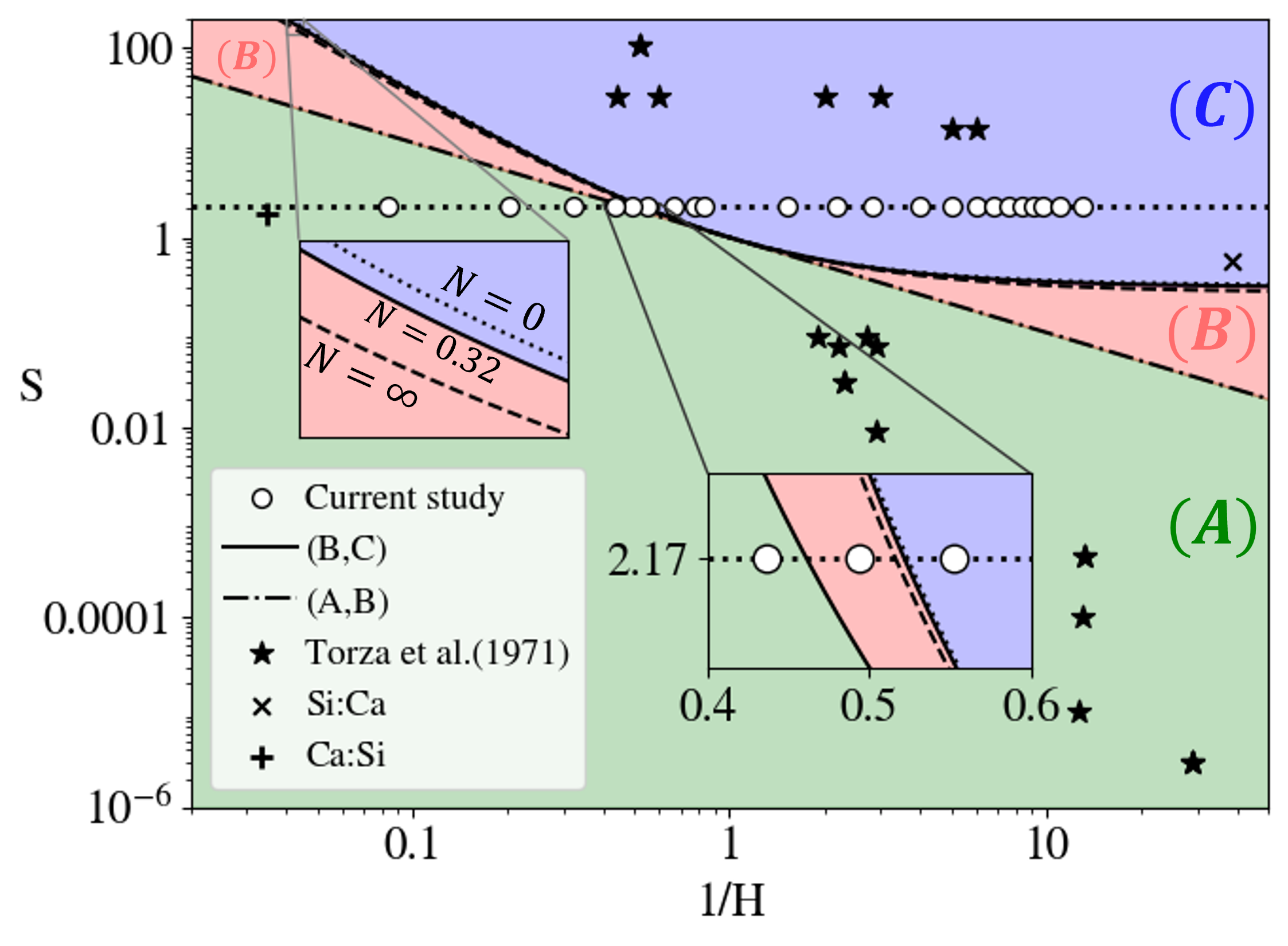}
\caption{{\bf Parameter space for experiments.} 
Phase diagram as a function of the dielectric constant ratio $H$ and the inverse conductivity ratio $1/H$. The separation between different droplet deformation regimes is provided by the $S = H$ (dash–dot line) threshold line and the 
$S = H (1 + (5S(1+N)(1/S - 1)^2)/(16 + 19N))$ (solid) threshold line. 
These two lines separate the three distinct drop deformation regimes defined in Table~\ref{tab:Classes} - Class A (green), Class B (red), and Class C (blue) - based on their predicted electrohydrodynamic behavior. 
The right inset highlights the narrow region corresponding to Class B, located between Classes A and C. 
The white circles display the samples made in this study, varying $1/H$. Systems from previous studies are also shown: silicone oil in castor oil (Si:Ca, $\times$), castor oil in silicone oil (Ca:Si, $+$), and
results from \cite{torza1971electrohydrodynamic}, indicated by {\bf $\star$}, which all include an aqueous phase, with inherent limits on electric field strength. 
The left inset demonstrates that varying the viscosity ratio $N$ from $0$ (dotted line) to $\infty$ (dashed line) 
has a negligible effect on the two threshold lines. Our experiments (white circles), provide a means to tune the system systematically across the distinct regions of this parameter space.
}
\label{fig:phaseSpace}
\end{figure*} 

Controlling conductivity by continuously varying the volume fraction (white circles in Figure~\ref{fig:phaseSpace} show sample conductivities used in this study) allows us to quantitatively probe the Torza equations~\cite{torza1971electrohydrodynamic}. 
For typical materials parameters such as those in this study, Class B is a very small part of the parameter space of the two important parameters, the dielectric constant ratio, $S$, and the conductivity ratio, $H$.  (Figure~\ref{fig:phaseSpace}). Class B occurs for $1/H = 0.046-0.052$ and has not been observed experimentally. Indeed, previous studies (indicated by {\bf $\star$}, $\times$ and $+$ in Figure~\ref{fig:phaseSpace}) are either deep in class A or class C. 
Mixing the miscible motor oil (Mo) and silicone oil (Si) as the continuous phase (see  SM Table I), we can keep the permittivity ratio $S \sim 2.7$ and the viscosity ratio $N \sim 5.9$ nearly fixed, while varying the conductivity ratio $H$ from 0.06 to 2.9 $via$
 careful control of the (Mo) volume fraction $\phi = (V_{Mo}/(V_{Si}+V_{Mo}))$ in the Mo-Si mixture. 
Varying $0.0 < \phi < 0.5$ results in $0.083 < 1/H < 12.98$.

Additionally, in the class C regime, tip-streaming and symmetry-breaking instabilities have been reported~\cite{brosseau2017streaming, vlahovska2019electrohydrodynamics}, and droplets can spontaneously rotate and acquire a steady tilted shape~\cite{quincke1896ueber, ha2000electrohydrodynamics, salipante2013electrohydrodynamic, he2013electrorotation}. With increasing field strength, droplets exhibit stable Quincke rotation, followed by irregular rotational dynamics~\cite{sato2006behavior, salipante2013electrohydrodynamic}. 
Collective multi-drop motions have been observed and categorized~\cite{varshney2012self}; convective and even chaotic motions at strong electric field occur in spite of the low Reynolds number (of order $\mathcal{O}(10^{-4})$) of the system. The dimensionless strength of EHD forcing is captured by the electric capillary number, $Ca_E$, which is the ratio of the electric (EHD) stress relative to
interfacial tension, and is expressed as
\begin{equation}
Ca_E \equiv 
f_2(H,S,N)
\frac{\varepsilon_0 \kappa_{1} a E^2}{\gamma_\mathrm{int}},
\label{CaE}
\end{equation}
where $f_2(H,S,N) = \frac{
    9\, \lvert S^{-1} H - 1 \rvert \, N^{-1}
}{
    10 (H + 2)^2 (N^{-1} + 1)
}.$


\section{Materials and Methods}
\subsection{Emulsion Composition and Preparation}

The emulsion consists of NOW Solutions castor oil, dyed with Nile Red (abbreviated as Ca), serving as the dispersed phase (phase 1). The continuous phase (phase 2) is a mixture of two miscible oils: Irving MAX1 10W-30 motor oil (abbreviated as Mo) and Dow Corning Dowsil 550 silicone oil (abbreviated as Si). The relevant properties are listed in Table~\ref{Table_properties}.
The two oils are mixed using a vortexer (Scientific Industries Vortex-Genie 2). The sample is then placed in an ultrasonic bath (Cole-Parmer 8850) for 5 minutes to remove air bubbles, followed by a 10-minute resting period to allow it to reach room temperature and for any surface foam to dissipate. The sample cell (described next) is filled by pipetting the emulsion and subsequently sealed using NOA 68 Optical Adhesive (Norland Products Inc.).
\begin{table*}[ht!]
\centering
\begin{ruledtabular}
\begin{tabular}{ccccc}
\textbf{Phase} & \textbf{Material} &
\textbf{Electrical Conductivity} &
\textbf{Rel. Permittivity} &
\textbf{Viscosity} \\
 &  &
$\sigma$ ($\mathrm{pS/cm}$) &
$\kappa$ ($\epsilon/\epsilon_0$) &
$\eta$ ($\mathrm{mPa{\cdot}s}$) \\
\hline
1 & Castor Oil {[}Ca{]} &
$3.96\pm0.01$ &
$5.6\pm1.3$ &
$819\pm7$ \\
\hline
2 & Motor Oil {[}Mo{]} &
$102.5\pm0.1$ &
$2.58\pm0.52$ &
$149\pm4$ \\
 & Silicone Oil {[}Si{]} &
$0.33\pm0.01$ &
$2.6\pm0.1$ &
$147\pm2$ \\
\end{tabular}
\end{ruledtabular}
\caption{Experimentally relevant materials parameters: electrical conductivity, relative dielectric permittivity, and viscosity for castor oil, motor oil and silicone oil. In the two-phase emulsion, fluid 1 ($\sigma_1$, $\epsilon_1$, $\eta_1$) is castor oil, while fluid 2 is a mixture of motor oil and silicone oil, with varying $\sigma_2$ but fixed $\epsilon_2$ and $\eta_2$.}
\label{Table_properties}
\end{table*}

A low conductivity is achieved by low volume fraction $\phi$ see SM Table I. Conductivity measurements were performed using a Scientifica conductivity meter (Model 627). The measurement setup involves immersing a probe, consisting of two parallel tubes, into the sample. 
The relative permittivity (\( \kappa \)) of three oils was measured using a rotational rheometer equipped with electrically isolated, Teflon-coated parallel-plate electrodes. An electric field was applied across the sample, and the resulting normal force was recorded. This force follows the parallel-plate capacitor relation, $F = \frac{\epsilon_0 \kappa A}{2d^2} V^2$, i.e., a linear relation between $F$ and $V^2$: the relative permittivity was obtained from the slope (details in SM Section (A)).

The interfacial tension $\gamma_{int}$ was obtained in a separate experiment from the time-dependent deformation of a drop under an electric field. 
This experiment yielded a characteristic time from which we extracted $\gamma_{int}$: see SM Section (A) for details. Viscosities were measured using an Anton Paar Physica MCR 301 rotational rheometer. 

As shown in Table~\ref{Table_properties}, Mo and Si have nearly identical viscosities and dielectric constants; thus, varying their volume fractions primarily affects the conductivity of the mixture without significantly altering other properties. For reference, the conductivity of distilled water is of order $\mu$S/cm, so while the conductivity of the continuous fluid can in principle be varied by a factor of 300, all measurements reported are nevertheless still small enough as to remain in the ``leaky dielectric'' regime. Moreover, in order to effectively probe the phase diagram in Figure~\ref{fig:phaseSpace}, we only varied the volume fraction from 0 to 0.5: see SM Table I for the experimental sample parameters.

\begin{figure}[htp!]
     \centering
   \includegraphics[width=0.3\textwidth]{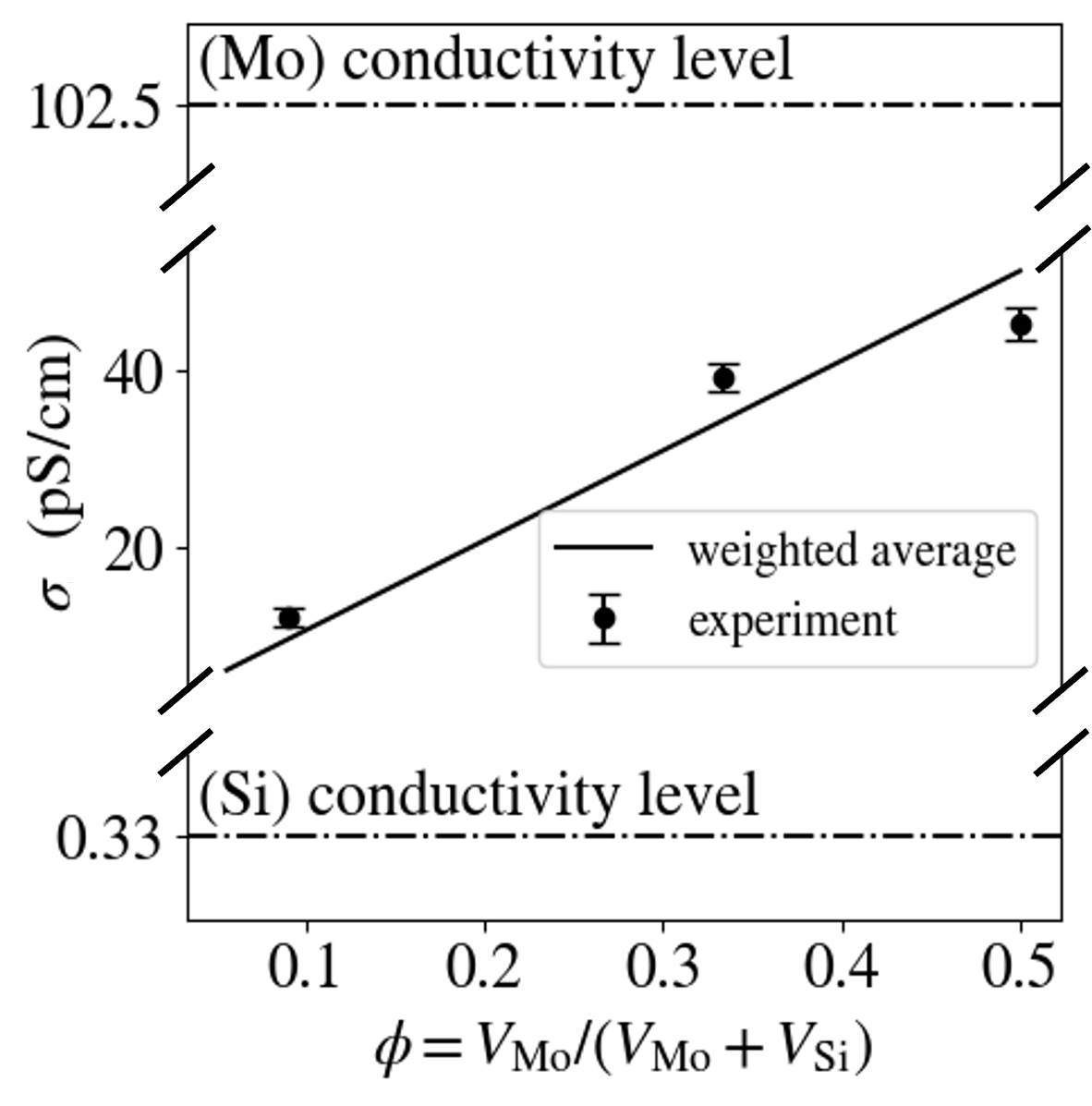}
        \caption{Linear variation of the inverse conductivity ratio $1/H$ with droplet volume fraction $\phi$. Conductivity values for intermediate compositions were estimated using this linear trend. Three representative samples prepared at different volume fractions used to confirm the relation.}
        \label{fig:mixing_sigma}
\end{figure}

The inverse conductivity ratio $1/H$ varied roughly linearly with volume fraction $\phi$. Shown in Figure~\ref{fig:mixing_sigma}
are three samples made at different volume fractions. Since our conductivity measurements required larger volumes (17 ml) for precise measurements, we used this linear dependence to obtain the conductivities for other prepared volume fractions.

\subsection{Sample cell geometry}
\label{Sec:VFC}
\begin{figure}[ht]
     \centering
        \includegraphics[width=0.3\textwidth]{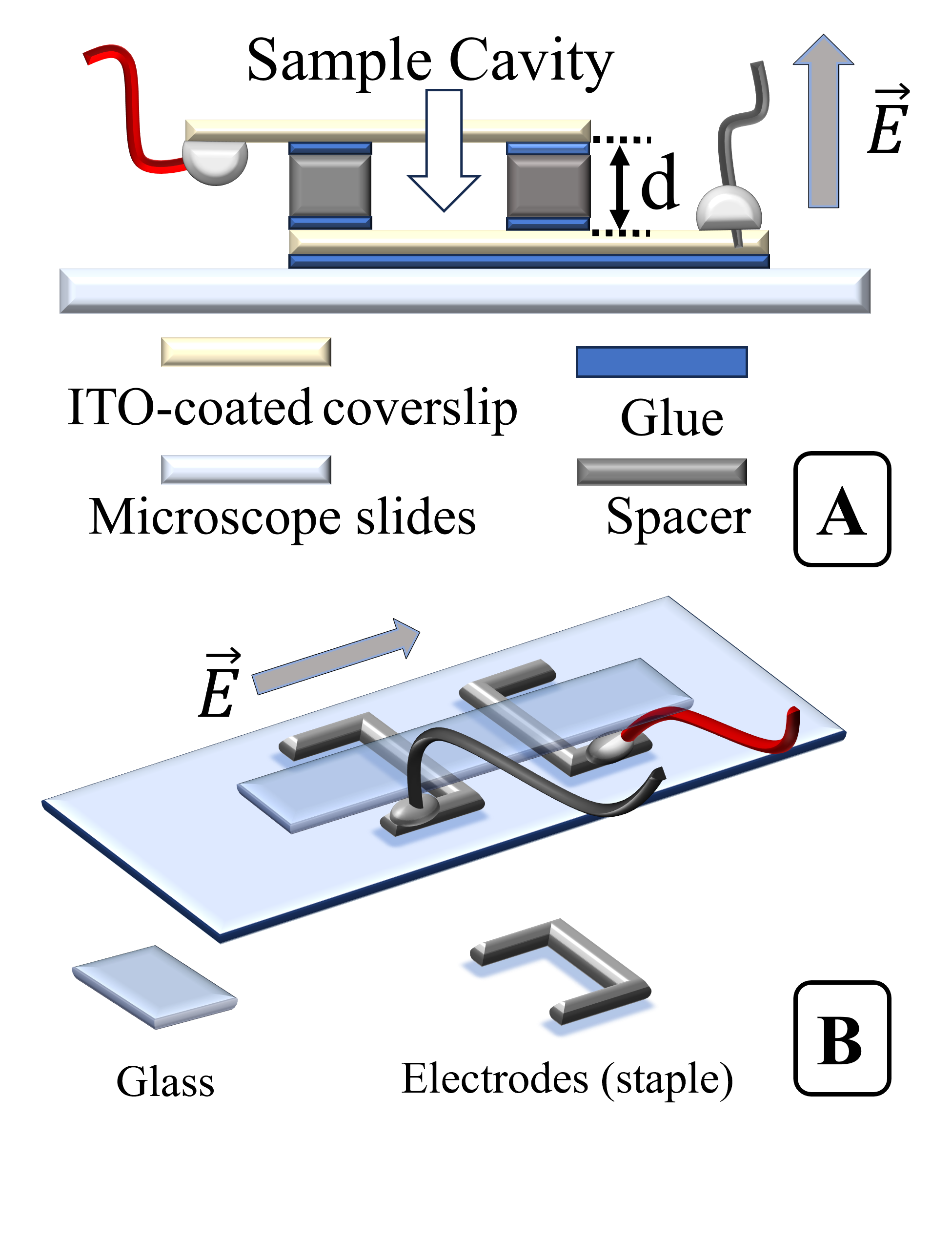}
        \caption{{\bf Sample cell geometry:} Schematic of (A) vertical-field and (B) horizontal-field cells.}
        \label{Fig:Cells}
\end{figure}

Two sample-cell geometries were used in this work. A ``vertical-field'' cell (Figure~\ref{Fig:Cells}(A)) consisted of plates coated with indium tin oxide (ITO) separated with two spacers on opposite sides of the bottom plate leaving the middle empty for filling with emulsion.
This type of cell is suitable for high-field studies and visualization of droplet flows. The lateral dimension of the cell was 1 cm $\times$ 1.5 cm; cell thickness $d$ was set at 200 $\mu m$. As demonstrated in Reference \cite{tadavani2016effect},  
such cells are effectively 3-dimensional for the droplets in a field with average drop size being much smaller than $d$.

On the other hand, ``horizontal-field'' cells are specifically designed for imaging drops from a viewing direction that is perpendicular perspective to the electrical field, as illustrated in Figure \ref{Fig:Cells}(B). This type of cell is suitable for sensitive measurements of steady-state and dynamical droplet shape (and from the former, the critical frequency $\nu_c$). Two staples are glued and cured to a microscopic slide, after which a piece of cover slip is affixed to the top of the electrodes.

\subsection{Image Processing and Data Analysis}
Image sequences are recorded using a Nikon Eclipse 80i upright microscope coupled with a PCO.edge 4.1 sCMOS camera. Bright-field images are captured with a 4x objective lens (NA = 0.13) at 160 frames per second and a resolution of 2048 $\times$ 1024 pixels. The images are later cropped into two 1024 $\times$ 1024 pixel regions. This approach takes advantage of the rolling shutter mechanism, allowing the acquisition of more data without compromising the frame rate.

Particle Image Velocimetry (PIV) \cite{abdulwahab2020review} and Differential Dynamic Microscopy (DDM) \cite{lattuada2025hitchhiker} are combined to analyze the image sequences. Velocity fields were computed using OpenPIV v0.25.3\cite{openpiv} and DDM analysis was performed using FastDDM v0.3.14 \cite{lattuada2025hitchhiker}. The flow field is characterized using the PIV method, implemented via the OpenPIV Python library, applied to the bright-field image sequences of the oil-in-oil emulsion, where the droplets serve as tracers providing information on the background flow. The total kinetic energy is defined based on the obtained velocity field:
 $E_{xy}(x, y, t) = (v_x^2(x, y, t) + v_y^2(x, y, t))/2$.
 
In two dimensions, particle motion can be purely diffusive, with the mean squared displacement (MSD) scaling as $\langle \Delta r^2(t) \rangle \sim 4Dt$, or it can exhibit anomalous diffusion, characterized by $\langle \Delta r^2(t) \rangle \sim K_\gamma t^{\gamma}$, where $K_\gamma$ is a generalized diffusion coefficient and $\gamma$ is the anomalous exponent: $\gamma < 1$ corresponds to subdiffusion, $\gamma = 1$ to normal diffusion, and $\gamma > 1$ to superdiffusion. In our system, the displacement distribution is Gaussian (see SM Section C). Unlike in equilibrium, however, the MSD follows a power-law scaling with $t$, suggesting fractional Brownian motion (FBM) as an appropriate model for droplet dynamics in the active regime. FBM~\cite{metzler2000random, metzler2004restaurant, metzler2014anomalous}, originally developed to describe anomalous diffusion under external forces~\cite{mandelbrot1968fractional}, represents a Gaussian self-similar process. In this framework, the MSD scales as $\langle \Delta r^2(t) \rangle \sim K_\gamma t^{\gamma}$.
Using the same frames, the motion of castor oil droplets in our emulsion is characterized by the DDM analysis.

In DDM, one analyzes the difference signal  
\[
D(x, y; \Delta t) = I(x, y; \Delta t) - I(x, y; 0),
\]  
where \( I(x, y; t) \) is the image intensity at position \( (x, y) \) and time \( t \). Taking the Fourier transform yields  
\begin{equation}
F_D(q_x, q_y; \Delta t) = \int D(x, y; \Delta t) \, e^{-i2\pi (q_x x + q_y y)} \, dxdy,
\end{equation}  
with corresponding intensity  
\begin{equation}
|F_D(q_x, q_y; \Delta t)|^2 = P_S = A(q)[1 - f_T(q, t)] + B(q).
\end{equation}  
The intermediate scattering function \( f_T(q, t) \) takes the form  
\begin{equation}
f_T(q, t) = \exp\left[-\frac{q^2}{4} \langle \Delta r^2(t) \rangle\right] = \exp\left[-K_\gamma q^2 t^\gamma\right],
\end{equation}  
providing a direct link between image fluctuations and anomalous dynamics. Defining the characteristic decay time as  
\begin{equation}
\frac{1}{\tau(q)} = (K_{\gamma} q^2)^{\frac{1}{\gamma}},
\label{eq:invTauq2}
\end{equation}  
we obtain  
\begin{equation}
P_S = A(q)\left[1 - e^{-t/\tau(q)}\right] + B(q).
\end{equation}  
Using the DDM data, one can fit values of \( \tau(q)^{-1} \) for different wavevectors. Taking the logarithm of equation \ref{eq:invTauq2} yields:  
\begin{equation}
\log\left(\frac{1}{\tau}\right) = \frac{1}{\gamma} \log(K_{\gamma}) + \frac{1}{\gamma} \log(q^2),
\label{mj}
\end{equation}  
Thus, \( \gamma \) can be interpreted as the inverse slope of a plot of \( \log(1/\tau) \) versus \( \log(q^2) \); these steps are summarized in Figure \ref{fig:DDM_steps}. 


\section{Results}
\subsection{Varying conductivity to validate theory}

We systematically varied the conductivity of the continuous phase by adjusting the mixing ratio of motor oil and silicone oil. To ensure the system initially belongs to class C, we began with a (low) inverse conductivity ratio of 0.78, corresponding to a motor oil volume fraction in fluid 2 (the continuous phase) of \( \phi = 0.027 \). 
The motor oil content was gradually increased up to a 0.5 fraction by mixing 3 $m L$ of each oil. The effective conductivity of each mixture was estimated using a weighted average of the individual conductivities. To validate this assumption, we directly measured the conductivity at three representative ratios—0.09, 0.33, and 0.5—and found good agreement with the expected weighted average trend (Figure~\ref{fig:mixing_sigma}). These measurements provide confidence in extending the interpolation across all tested compositions.
\begin{figure*}[th!]
\centering
\includegraphics[width=0.85\textwidth]{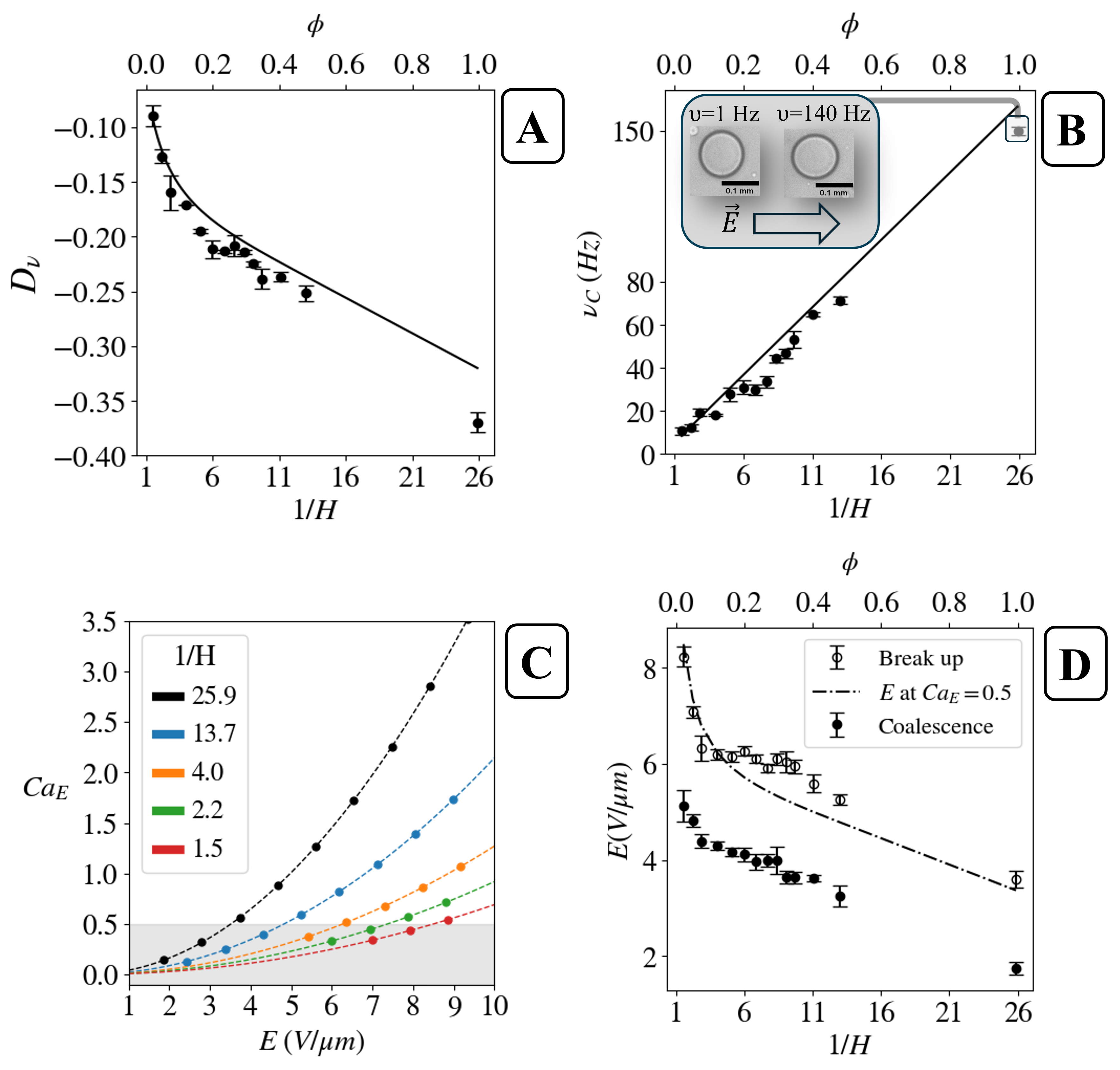}
\caption{{\bf Varying conductivity to validate theory:} 
(A) Deformation parameter \( D_{\nu} \) for dc fields, $\nu=0$ as a function of inverse conductivity ratios \( 1/H \) at $E=2 V\mu m^{-1}$. The theoretical curve (with all parameters known, so is not a fit) employs Eq.~\ref{eq:deformation} and Eq.~\ref{eq:discriminationFunction}. (B)  The critical frequency, \(\nu_c \), where drops switch from oblate to prolate, increases linearly as a function of \( 1/H \). Inset: Castor oil droplet in pure motor oil under a horizontally applied a.c. electric field. The electric field is horizontal in the image. The threshold frequency for the left drop in the panel ($1/H=1.5$) to be spherical in a $E = 3.5~V/\mu m $ field was $\nu_c = 1$~Hz, while it was 140~Hz for the right drop ($1/H=25.9$). The (parameter-free) theory curves in (A) and (B) employ parameters (see also Table \ref{Table_properties}) $S=2.17$, $N=5.49$, $a \approx 30 \mu m$ and $\gamma_{\mathrm{int}}$ (values are presented in Table \ref{Table_properties}). (C) Electric capillary number $Ca_E$ (Eq.~\ref{CaE}) for droplets of radius $30\,\mu\mathrm{m}$ plotted versus electric field for different inverse conductivities $1/H$. 
(D) Threshold fields for coalescence and breakup both decrease as a function of increasing \( 1/H \). These observations go beyond small-deformation theory. The dash-dot line is obtained by solving Eq.~\ref{CaE} for $E$ at $Ca_E = 0.5$, providing a theoretical threshold that lies within the experimentally observed range.}
\label{fig:conductivity_composite}
\end{figure*}

Figure~\ref{fig:conductivity_composite}(A) shows the deformation parameter \( D_{\nu} \) at \( E = 2\ \mathrm{V}/\mu\mathrm{m} \) as a function of the inverse conductivity ratio \( 1/H \). Each data point corresponds to a specific \(\phi \). Negative values of \( D_{\nu} \) indicate oblate deformation. The experimental results are consistent with the theoretical prediction (with no fitting parameters!) and demonstrate that increasing \( 1/H \) leads to stronger deformation, quantitatively as expected from electrohydrodynamic theory.
In Figure~\ref{fig:conductivity_composite}(B), the critical frequency \( \nu_c \), marking the transition from oblate to prolate shapes under an AC field, is plotted against \( 1/H \). The data show excellent agreement with the theoretical prediction across the entire conductivity range. Again, it is worth noting that there are no free parameters in the theory predictions for \( D_{\nu} \) and \( \nu_c \), so the quantitative agreement is excellent. This consistency supports the validity of the electrohydrodynamic model, at least at small deformations, and confirms that conductivity tuning enables systematic control of droplet behavior.

Figure~\ref{fig:conductivity_composite}(C) plots Eq.~\ref{CaE} for the calculated electric capillary number as a function of applied electric field for different inverse conductivities $1/H$ (for a mean drop radius $a=30\,\mu\mathrm{m}$). The purpose of this plot is to show that, e.g. at $Ca_E = 0.5$, the electric field threshold for drop breakup increases as $1/H$ decreases from 25.9 (corresponding to a motor oil continuous medium~\cite{bahraminasr2026electrorheoimaging}) to 1.5 (corresponding to a silicone oil continuous medium).

Experimentally, for each sample (at a given $1/H$), as one increases the electric field, one first sees drop motions. There is a threshold field, recorded by visual observation, where neighboring drops begin coalescing (Figure~\ref{fig:conductivity_composite}(D), solid circles). At a higher field threshold, drops begin to break up. For a given mean drop size (around 30 $\mu$m radius, generated by initial breakup under a high field of \( 10\ \mathrm{V}/\mu\mathrm{m} \)), there is a second threshold field (also observed visually) where drops begin to break up. The threshold electric field for coalescence and for breakup (open circles) is shown in Figure~\ref{fig:conductivity_composite}(D). As \( 1/H \) increases, both breakup and  coalescence thresholds (open and filled circles in Figure~\ref{fig:conductivity_composite}(D)) decrease markedly: solving Eq.~\ref{CaE} for $E$ at $Ca_E = 0.5$, one obtains electric field strengths (dash-dot line) that compare well with experimental break-up theshold values (open symbols).

This trend can be understood in terms of the charge relaxation time, $t_{1,2} = \epsilon_{1,2}/\sigma_{1,2}$, which characterizes the time required for free charges in fluid 1 and 2 to accumulate at the interface: note that the parameter ratio $S/H = t_1/t_2$. In our system, we increased $1/H$ keeping $S$ constant, and the ratio $t_1/t_2$ was varied from 0.2 to 60. When $1 < t_1/t_2 < 60$, the outer fluid (continuous phase) supplies charge more rapidly to the interface, meaning that interfacial charge accumulation is primarily governed by the properties of the continuous phase. This behavior directly influences both breakup and coalescence. Enhanced charge accumulation increases interfacial electric stresses, which facilitates droplet breakup at lower field strengths. Coalescence occurs when two oppositely induced-charged droplets align and attract each other\cite{tadavani2016effect}.

Finally, the excellent agreement between experimental measurements and theoretical predictions for both the deformation parameter \( D_{\nu} \) and the critical frequency \( \nu_c \) confirms the predictive power of the Torza theory~\cite{torza1971electrohydrodynamic}. We have also obtained threshold fields which are not obtainable from a linear (small-deformation) theory in a system where one can vary the ratio of charge relaxation times $t_1/t_2$ without affecting \emph{any other parameters}. From a practical standpoint, while the qualitative sequence of behaviors remains similar across different conductivity contrasts, the critical field thresholds {decrease significantly, enhancing the utility of this phenomenon, and providing greater access to the physics of the high-field regime for class C droplets.

\subsection{PIV and the spatio-temporal energy spectrum}

For castor oil in pure motor oil (Ca:Mo, with \(1/H = 25.9\), the system is in the class C regime. As the electric field increases from $E = 0$ to $3.7~\mathrm{V}\,\mu\mathrm{m}^{-1}$, the system transitions from a regime characterized by droplet coalescence to one dominated by the violent breakup of large droplets into smaller ones (SM Movie 1), accompanied by intensified droplet motion and unsteady flows. The strength of the motion grows significantly with increasing field. 

The flow fields were quantified using particle image velocimetry (PIV). 
Figures~\ref{fig:PIV_panel} (A) and (B) show representative velocity fields at \(E = 3.7\) and \(9.3~\mathrm{V}\,\mu\mathrm{m}^{-1}\), respectively. As the applied field is increased from \(E = 0\), the system undergoes a transition from a regime dominated by droplet coalescence to one characterized by the breakup of large droplets into smaller ones (SM Movie 1). In this breakup-dominated regime, the motion of droplets becomes markedly more vigorous, and the velocity fields reveal increasingly complex and unsteady flows.

\begin{figure*}[ht!]
\centering
\includegraphics[width=0.7\textwidth]{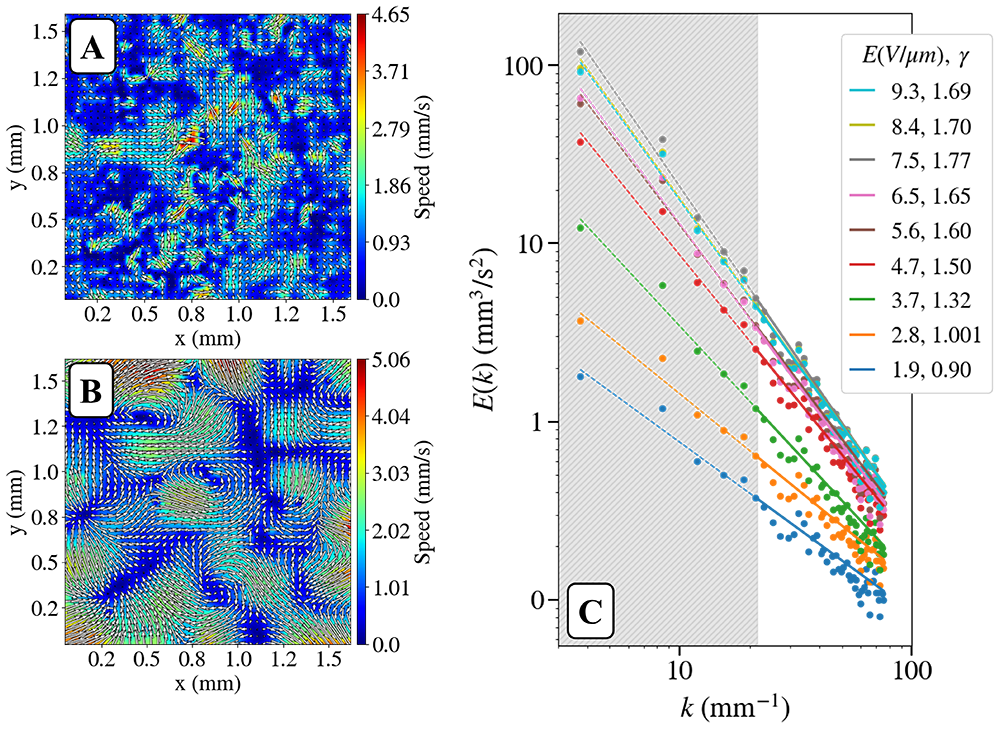}
\caption{{\bf PIV and the spatio-temporal energy spectrum:} castor oil in pure motor oil system ($1/H = 25.9$): (A)–(B) Velocity fields at two different field strengths, E = $3.7 V\mu m$ and $9.3 V\mu m$, showing increased complexity at higher $E$. (C) Power spectra exhibiting power-law behavior with slopes approaching Kolmogorov scaling.
The shaded region denotes $k$ values that correspond to lengthscales larger than the sample thickness: with decreasing $k$ the flow becomes increasingly 2-dimensional and deviations from scaling are expected.}
\label{fig:PIV_panel}
\end{figure*} 

From these PIV datasets, the spatio–temporal energy spectrum \(E(k)\) was computed [Fig.~\ref{fig:PIV_panel}(C)]. The spectra exhibit a power-law dependence, \(E(k) \sim k^{-\alpha_k}\), with fitted exponents increasing with field strength from \(\alpha_k = 0.91 \pm 0.08\) to \(\alpha_k = 1.75 \pm 0.12\). In Fig.~\ref{fig:PIV_panel}(C) we highlight in grey the wavenumber region corresponding to 
lengthscales exceeding the sample thickness, where flow should transition toward effectively two-dimensional behavior. Consequently, we do not expect reliable scaling at smaller 
$k$. Notably, for the highest fields and over a range of wavenumber, the scaling exponent is experimentally consistent with the Kolmogorov value of \(\alpha_k = 5/3 = 1.67\), characteristic of an inertial energy cascade in fully developed turbulence. Similar $5/3$ scaling has been reported in active-matter turbulence \cite{bhattacharjee2022activity, bourgoin2020kolmogorovian}. At low wavenumbers, deviations from this scaling (i.e., systematic deviation of symbols from the dashed lines) mark the breakdown of self-similar behavior, and simply reflect the influence of finite system size.

\subsection{Rheology}
Macroscopic rheological measurements reveal the time evolution of electric field–induced shear stresses, recorded at the top electrode. Figure~\ref{fig:rheo}(A) presents the time series of shear stress fluctuations, defined as $\tau - \langle \tau \rangle$, where $\langle \cdot \rangle$ denotes a time average, for various values of $E$ (each indicated by a different color). These fluctuations are identified as the Reynolds shear stress, representing the contribution of chaotic motions to the mean shear stress. The data clearly show that the internal flow becomes increasingly unsteady with rising electric field strength, as reflected by the growing amplitude and complexity of the stress fluctuations.
Figure~\ref{fig:rheo}(B) displays the power spectrum of the time series of Reynolds shear stress ($S_f$) for selected values of $E$. The power law exponents are consistent with those obtained from microscopic imaging. For the emulsion of castor oil in pure motor oil at $E = 3.7~\mathrm{V}/\mu\mathrm{m}$, the spectrum exhibits power-law behavior with an exponent of $S(\nu) \sim \alpha_{\nu}$ of approximately $1.3$, while at $E = 9.3~\mathrm{V}/\mu\mathrm{m}$}, the exponent approaches $5/3$, coinciding with the value for fully developed inertial-like turbulence. 
\begin{figure}[ht!]
\centering
    \includegraphics[width=0.75\linewidth]{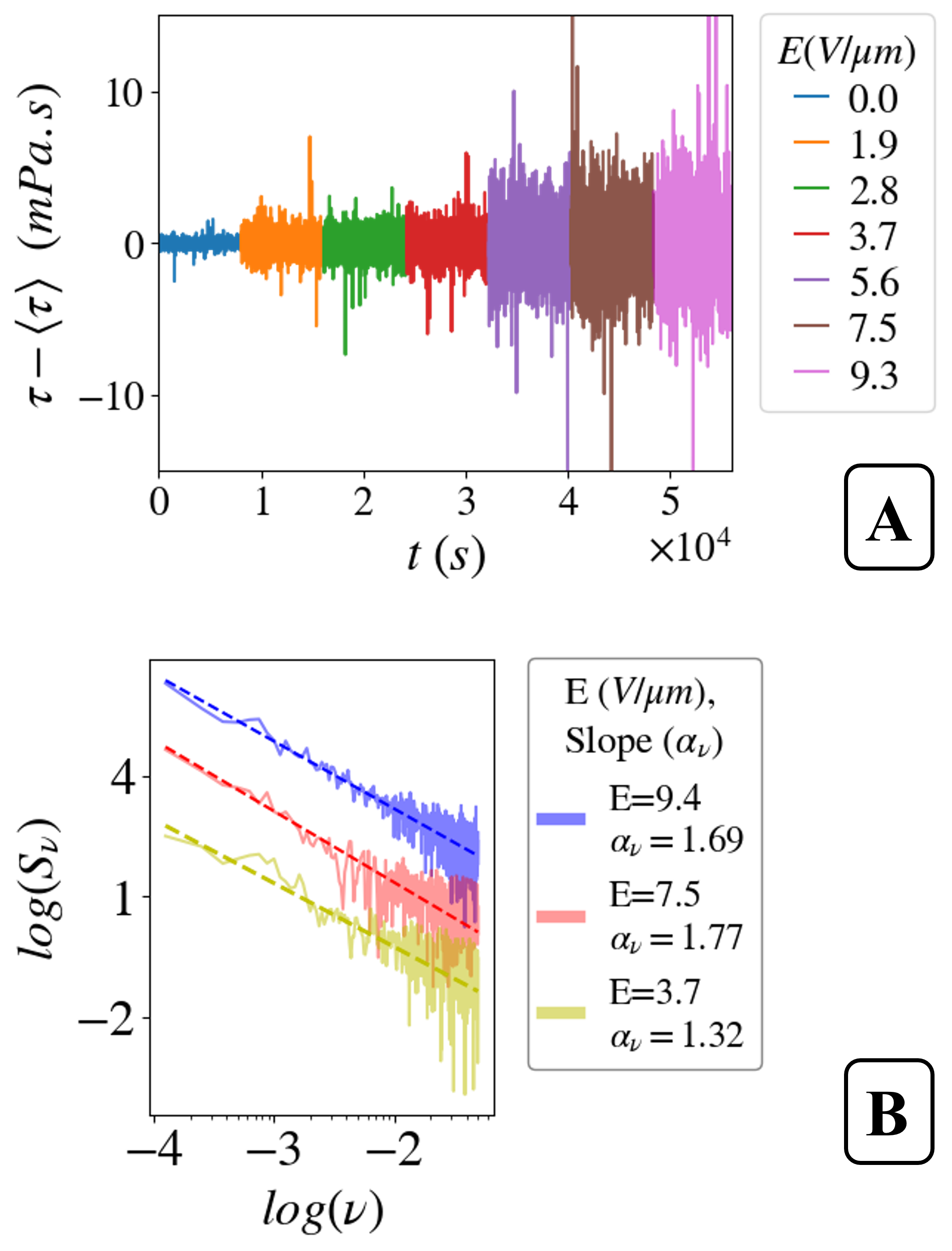}
\caption{\textbf{Electric-field-driven shear stress fluctuations and spectra} (A) Castor oil in pure motor oil (\(1/H = 25.9\)): Temporal fluctuations of shear stress, $\tau - \langle \tau \rangle$, with increasing electric field $E$. (B) Castor oil and pure motor oil system: Power spectrum of the time series of Reynolds shear stress ($S_f$).}
\label{fig:rheo}
\end{figure}

\begin{figure}[ht!]
\centering
    \includegraphics[width=1\linewidth]{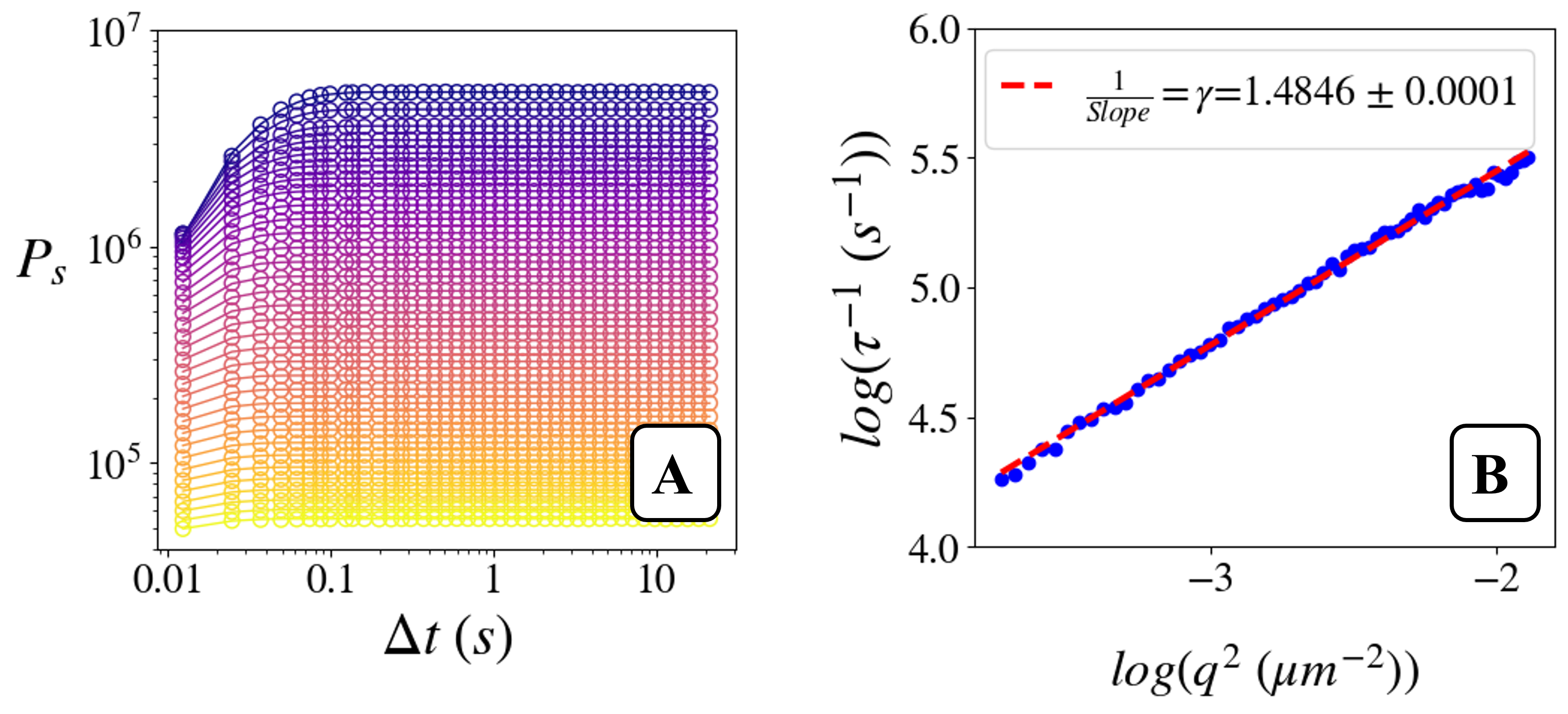}
\caption{\textbf{Differential Dynamic Microscopy (DDM) of active motions}. (A) Sequential microscopy frames $I(x,y;t)$ are Fourier transformed and differenced, $\Delta S = \mathscr{F}(I(x,y;\Delta t)) - \mathscr{F}(I(x,y;0))$, to compute the structure function $P_S(q,\Delta t) = |\Delta S|^2$, which is shown as a function of $\Delta t$. 
Each $P_S(q,\Delta t)$ fitted to a function with a form proportional to $e^{(-\Delta t/\tau(q))}$ where $\tau(q)$ is relaxation time. (B) Plotting $\log(1/\tau)$ versus $\log(q^2)$ reveals a power-law dependence with slope $\gamma$, the anomalous diffusion exponent characterizing the underlying transport dynamics. The sample shown here, corresponding to castor oil in pure motor oil system ($1/H = 25.9$), exhibits active, super-difusive droplet motions consistent with a $\gamma = 1.5$ power law.}
\label{fig:DDM_steps}
\end{figure}

\subsection{Dynamic difference microscopy (DDM) of active motions}
In active systems, the mean squared displacement (MSD) of tracer particles provides crucial insight into the underlying flow dynamics. While passive, equilibrium systems typically exhibit linear scaling, $\langle \Delta x^2(t) \rangle \sim t$, characteristic of classical diffusion, active matter often displays anomalous transport with nonlinear scaling. Recent theoretical and numerical work~\cite{mukherjee2021anomalous} has shown that in dense active suspensions, the MSD follows a super-diffusive form, $\langle \Delta x^2(t) \rangle \sim t^{\gamma}$, with $1 < \gamma < 2$, reflecting enhanced transport due to self-propulsion and collective interactions.

In our system, tracking individual droplets proves impractical for long periods of time. However, DDM-derived MSD exponents $\gamma$ (Figure~\ref{fig:exponent}(B)) reveal that for all applied field strengths, we find $\gamma > 1$, indicating super-diffusive behavior. Briefly, we begin with raw microscopy image frames $I(x,y;t)$. We then use the difference of the Fourier transform of frames ($\Delta S = \mathscr{F}(I(x,y;\Delta t)) - \mathscr{F}(I(x,y;0))$) to calculate the structure function $P_S(q,t) = |\Delta S|^2$ (Figure \ref{fig:DDM_steps}(A)). The decay of $P_S(q,\Delta t)$ is fitted to the function $e^{(-\Delta t/\tau(q))}$ to obtain a q-dependent relaxation time $\tau(q)$. By plotting $\log(\tau^{-1})$ against $\log(q^2)$ (Figure \ref{fig:DDM_steps}(B)), we extract $\gamma$, anomalous diffusion exponent. The key assumption here is the Gaussianity of MSD \cite{berne2000dynamic}. The validity of this assumption has been verified in SM Section (C).

\subsection{Multiple views of the transition to turbulent flows}

Varying the conductivity of the continuous phase reveals a notable influence on the energy spectrum exponent $\alpha_k$. While the emergence of a power-law spectrum persists across all conductivities, the transition into the active regime shifts to higher electric field strengths as the conductivity ratio $1/H$  decreases. This reflects the reduced efficiency of charge accumulation at the interface in low-conductivity systems, delaying the onset of strong electrohydrodynamic flows. 
\begin{figure*}[ht!]
       \centering
        \includegraphics[width=1\textwidth]{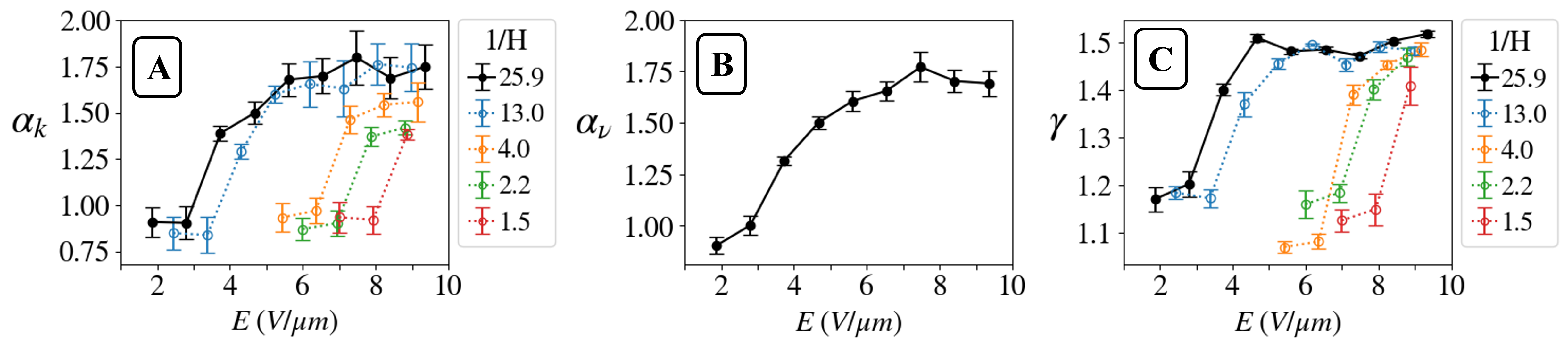}
    \caption{ {\bf Multiple views of the transition to turbulent flows:}
    Comparison of power-law exponents of (A) $E(k)$ vs. $k$, ($\alpha_k$, from Figure~\ref{fig:PIV_panel}), (B) $E(\nu)$ vs. $\nu$, ($\alpha_{\nu}$, from Figure~\ref{fig:rheo}) and (C) the MSD vs $t$, ($\gamma$, from Figure~\ref{fig:DDM_steps}), as functions of field strength for different inverse conductivity ratios $1/H$: all three reach a plateau above $E \sim 5 \mathrm{V}/\mu\mathrm{m}$ for $1/H = 25.9$. The threshold field for the plateau increases with decreasing $1/H$.}
    \label{fig:exponent}
\end{figure*}
For example, as shown in Figure~\ref{fig:exponent}(A), when the inverse conductivity ratio \(1/H\) is $25.9$, a clear $5/3$ spectral plateau is observed at $E = 5.6~\mathrm{V}/\mu\mathrm{m}$. In contrast, for a lower inverse conductivity ratio of $11.6$, even at a higher field of $E = 9.3~\mathrm{V}/\mu\mathrm{m}$, the spectrum only reaches an exponent of $1.38 \pm 0.03$, indicating incomplete development of the inertial-like cascade. This explains the observation of a lower ($\alpha_k \sim 1.39 \pm 0.02$) power law in previous work~\cite{Varshney2016}, and shows that high conductivity ratio is essential to achieve fully developed multiscale flows.

The influence of conductivity on $\gamma$ mirrors that seen in the spectral exponent $\alpha_k$: a reduction in bulk conductivity systematically delays the onset of the super-diffusion plateau, shifting the field threshold for fully developed turbulence. For instance, as shown in Figure~\ref{fig:exponent}(B), at $E = 7.5 \,\mathrm{V/\mu m}$, $\gamma \sim 1.1 - 1.2$ when $1/H= 2.8 - 8.3$,
but rises to 1.5 when $1/H \sim 25$. Plotting $\alpha_k$ vs $\alpha_{\nu}$ shows a linear trend with a slope of 1, a strong indication that we are probing the same phenomenon with two different methods (SM Figure 3(A)). 
A previous particle-tracking EHD study in emulsions reported~\cite{tadavani2016effect} an MSD exponent $\gamma = 1.4 \pm 0.1$ and power-law scaling of the energy spectra~\cite{Varshney2016} with $\alpha_k$ and $\alpha_{\nu}=1.4$. 
Similarly, our conductivity-tuned system passes through ($\alpha_k=1.4$, $\gamma=1.4$) (SM Figure 3(B)) before reaching the plateau value of ($\alpha_k=5/3$, $\gamma=3/2$) once the field exceeds a conductivity-dependent threshold. $\gamma = 3/2$ has been observed in active motions in bacterial suspensions~\cite{Caspi2000, Hoffman2006, Kulic2008}. This correlation between $\gamma = 3/2$ and $\alpha_k = 5/3$ is intriguing. 
This agreement between bulk rheological measurements and microscopic analysis confirms the robustness of the observed electrohydrodynamic turbulence across scales. Finally, given the observed plateau behaviour in all exponents, we average over all exponents in the plateau regime for the pure (Ca:Mo) system to get more robust values: $\gamma = 1.494 \pm 0.003$, $\alpha_k = 1.73 \pm 0.06$, and $\alpha_{\nu} = 1.70 \pm 0.03$.
\section{Discussion}
Spontaneous chaotic flows at low Reynolds numbers have been observed in biology,  e.g., in bacterial suspensions \cite{dombrowski2004self, cisneros2007fluid,sokolov2007concentration,ishikawa2011energy,sokolov2012physical,wensink2012meso,dunkel2013fluid,patteson2018propagation,li2019data,xie2022activity}. 
At high concentrations, the collective motions of bacteria can give rise to instabilities with phenomena that are analogous to inertial turbulence \cite{alert2022active}. In electrically driven emulsions, the applied electric field and the associated charge transport play an analogous role, even showing pairs forming tandem structures \cite{sorgentone2022tandem} or displaying 
predator–prey-like dynamics due to non-reciprocal electrohydrodynamic interactions \cite{meredith2020predator}. However, the collective EHD response of emulsions has been explored only to a limited extent \cite{varshney2012self, varshney2014large, varshney2016multiscale, tadavani2016effect, tadavani2018anomalous, raju2021diversity, kach2023nonequilibrium, bahraminasr2026electrorheoimaging}.

As discussed in a recent review~\cite{sreenivasan2025turbulence}, turbulence should not be viewed as a single problem to be solved, but rather as a hierarchy of interconnected far-from-equilibrium behaviors that include energy injection, energy transfer across scales and dissipation. In addition to classical inertial turbulence controlled by the Reynolds number, other forms of multiscale flows have been uncovered: convective turbulence \cite{wu1990frequency}, elastic turbulence in polymer solutions \cite{steinberg2021elastic}, multiscale EHD flows~\cite{Varshney2016} and turbulence in bacterial suspensions and active matter \cite{alert2022active}. One can thus view turbulence, more broadly, as a strong coupling problem in far-from-equilibrium matter. 

In this work, we introduce a unique model system to probe electrohydrodynamics in a binary oil-in-oil emulsion, with control provided by the electric field strength, and tunability introduced \textit{via} the electrical conductivity in the continuous phase. By systematically varying the mixture ratio of silicone oil and motor oil in the continuous phase, we tuned the bulk conductivity while keeping viscosity and permittivity nearly constant, thus isolating the role of charge transport dynamics in controlling interfacial behavior. This allowed us to experimentally span the parameter space by varying the conductivity ratio $1/H$, and we demonstrate quantitative agreement with theory in the low-field regime. We also demonstrate control, and significant reduction with increasing $1/H$, of the transition threshold from low-electric-field drop deformations to high-electric-field unstable EHD behaviour.

We make the case that the multiscale dynamics observed in EHD emulsions are strongly analogous to fluid turbulence by employing three independent experimental techniques to uncover the underlying dynamical power laws. All measurements were repeated three times under identical conditions, and the reported values represent averages over these independent replicates, ensuring reproducibility and statistical reliability. While in the imaging modalities the droplets serve as tracer particles, rheometry directly probes stress fluctuations. A simulation study of the strong-field limit of these emulsions, while challenging, is currently in progress and is expected to provide further physical insight into the underlying mechanisms.

At high electric fields, microscopic imaging combined with particle image velocimetry (PIV) revealed the velocity fields and spectral signatures of these transitions. In the active regime, we observed robust power-law energy spectra, with exponents reaching $\alpha_k = 5/3$, identical to the classic Kolmogorov scaling for inertial turbulence. Previous reports of multiscale EHD flows in emulsions~\cite{Varshney2016} with a $\alpha_k \sim 1.4$ power law (similar to convective turbulence) were experiments conducted in a system with low $1/H$. We now see that the observed power law is not yet in the plateau region, i.e., not yet fully developed turbulence. Notably, the threshold for entering this turbulent regime shifted to higher electric fields as the conductivity decreased, reflecting a reduction in interfacial charge accumulation due to longer relaxation times in low-conductivity mixtures. Additionally, macroscopic rheometry measurements of the fluctuating shear stress also yielded a temporal power law scaling $\alpha_{\nu} = 5/3$, thus closely matching those obtained from microscopic velocity fields. 
Simultaneous measurements of the mean squared displacement (MSD) using differential dynamic microscopy (DDM) showed super-diffusive behavior across all conditions, with an exponent plateauing at the value $\gamma = 3/2$ for high enough fields and at the higher conductivity ratios. 
We show in this work that this power law is tunable with conductivity control which makes the strong-EHD-interaction limit experimentally accessible.

Even at the highest recorded droplet speeds of 5 mm/s, the Reynolds number for this system $Re \sim 0.01$.  
Despite the low Reynolds number, the flow exhibits the key features of turbulence sustained by non-inertial forces. In many active systems—such as bacterial suspensions~\cite{wensink2012meso,dunkel2013fluid} or driven colloids~\cite{bechinger2016active}—displacement statistics show non-Gaussian, Lévy-like tails. In contrast, droplet displacements in our emulsion remain Gaussian (SM Section (C)), even while the mean-squared displacement is super-diffusive. This combination rules out heterogeneous propulsion and indicates a statistically homogeneous, continuum-like turbulent state driven by collective electrohydrodynamic interactions among droplets. Similar Gaussian yet super-diffusive dynamics have been observed in confined active nematics~\cite{guillamat2017taming}, suggesting that we have an active turbulent regime sustained by EHD flows rather than by self-propulsion. Finally, our measurements show that the onset of chaotic, turbulence-like dynamics can be anticipated by examining the electric capillary number $Ca_E$, which quantifies the ratio of electric to interfacial stresses (Eq.~\ref{CaE}). In particular, across all conductivity ratios, the electric field at which the energy-spectrum and MSD exponents saturate into their plateau values coincides with $Ca_E \simeq 0.5$ (calculated using a mean droplet radius $a=30\,\mu\mathrm{m}$).

Together, these findings provide a comprehensive, multiscale view of how conductivity modulates the physical mechanisms underpinning flow transitions in emulsions subjected to electric fields. Beyond enriching our understanding of electrohydrodynamic instabilities, this work introduces conductivity as a precise and tunable experimental handle to engineer droplet behavior from the control of droplet coalescence to the generation of sustained active turbulence, while at the same time significantly reducing electric fields required for the  observation of this behaviour. These insights offer potential applications in microfluidic mixing, emulsion processing, lab-on-chip devices, and active material design where electric fields and droplet interactions play a central role.

\begin{acknowledgments}
The authors acknowledge the support of the National Science and Engineering Research Council of Canada (RGPIN-2019-04970 and RGPIN-2025-05494). This research was undertaken, in part, thanks to funding from the Canada Research Chairs Program.
\end{acknowledgments}

\section*{Data Availability Statement}
The data that support the findings of this study are available within the article and its supplementary material.

\bibliography{aipsamp}

\end{document}